\documentclass[10pt]{article}
\usepackage{amssymb}

\begin{document}

\markboth{Ladislav Hlavat\'y, Jan Vysok\'y} {Manin supertriples and
Drinfel'd superdoubles in low dimensions}

\title{Manin supertriples and Drinfel'd superdoubles in low
dimensions}

\author{Ladislav Hlavat\'y and Jan Vysok\'y
\thanks{E--mail: hlavaty@fjfi.cvut.cz, vysokjan@fjfi.cvut.cz}
\\{\it Faculty of Nuclear Sciences and Physical
Engineering,}
\\{ \it Czech Technical University in Prague,}
\\ {\it B\v rehov\'a 7, 115 19 Prague 1, Czech Republic}}
\maketitle

\begin{abstract}
Defining the real Lie superalgebra as real $Z_2$--graded vector
space we classify real Manin supertriples and Drinfel'd superdoubles
of superdimensions $(2,2),\ (4,2)$ and $(2,4)$. They can be used for
construction of  $\sigma$--models on supergroups related by
Poisson--Lie T--plurality.
\end{abstract}



\newtheorem{theorem}{Theorem}

\def\cf{{\mathcal {F}}}
\def\cd{(S|\tilde S)}
\def\cl{{\cal L}}
\def\cg{{\cal S}}
\def\tcg{{\tilde{\cal S}}}
\def\dpm{{\partial_\pm}}
\def\dmp{{\partial_\mp}}
\def\dxp{{\partial_+}}
\def\dxm{{\partial_-}}
\def\ttil{\tilde{X}}
\def\htil{\tilde{h}}
\def\that{\widehat{X}}
\def\ghat{\hat{g}}

\def\wh{\widehat}

\def\wt{\widetilde}

\def\sm{$\sigma$--model}
\def\sa{superalgebra}
\def\cfn{classification}
\def\tfn{transformation}
\def\eqn{equation}
\def\mtx{matrix}
\def\pltp{Poisson--Lie T--pluralit}
\def\pltd{Poisson--Lie T--dualit}

\def\dd{Drinfel'd double}
\def\dsd{Drinfel'd superdouble}
\def\mt{Manin triple}
\def\mst{Manin supertriple}

\def\half{\frac{1}{2}}\def\unit{{\bf 1}}
\def\real{{\bf R}}\def\complex{{\bf C}}

\def\cD{{\cal {D}}}\def\cS{{\cal S}}\def\cF{{\cal F}}
\def\bilform{$\langle.,.\rangle$}
\def\cmt{{\cal C}}

\def\Anj{A_{01}}\def\Ajn{A_{10}}\def\Ajj{A_{11}}
\def\Adn{A_{20}}\def\And{A_{02}}\def\Ajd{A_{12}}\def\Adj{A_{21}}\def\Adj{A_{21}}
\def\Njj{N_{11}}\def\Njjep{N_{11}^{\ \epsilon}}
\def\Ndjep{N_{21}^{\epsilon}}\def\Ndj{N_{21}}\def\Njd{N_{12}}\def\Njdep{N^\epsilon_{12}}
\def\Sjj{S_{11}}\def\Sdj{S_{21}}\def\Sdn{S_{20}}
\def\cjp{C^1_{p}}\def\cjn{C^1_{0}}\def\chalf{F}\def\cjhalfep{C^{1,\epsilon}_{p=-\half}}

\def\cdp{C^2_{p}}\def\cdn{C^2_{0}}\def\cdj{C^2_{1}}
\def\ct{C^3}\def\cc{C^4}\def\cpp{C^5_p}\def\cpn{C^5_0}
\def\Nabg{N_{12}^{\alpha,\beta,\gamma}}

\def\tfj{\wt{f^1}}\def\tbj{\wt{b^1}}\def\tf2{\wt{f^2}}\def\tb2{\wt{b^2}}

\section{Introduction} \label{Introduction}
\mt s and \dd s as well as their supersymmetric counterparts are
used for several purposes, e.g., for construction of solution of the
Yang--Baxter equations and their supersymmetric analogs
\cite{leiserg:solcyb,karaa:rmat}. Because of that, many cases of \mt
s, especially for simple \sa s were constructed.

Further investigation and classification of Manin triples and Drinfel'd doubles was boosted by discovery of
Poisson--Lie T--duality of \sm s by Klim\v{c}\'{\i}k and \v{S}evera\cite{klse:dna}. The low dimensional cases became a
convenient laboratory for investigation of the \pltd y and its extension to the concept of \pltp
y\cite{klse:dna,unge:pltp}.

As \sm s, where the \pltp y is used, are related to the string
theory it is natural to extend the \pltp y to the supersymmetric
structures either on the source space or in the target space. Useful
tool for understanding features of the \pltp y of \sm s with
supersymmetric target spaces is construction of simple and
potentially solvable cases. The first step for this is to find and
classify the low dimensional \mst s and \dsd s.

Several separate cases of low dimensional \mst s or, equivalently,
superbialgebras were investigated  in \cite{jusob:e2osp12} and an
attempt to classify the \mst s was done in \cite{ehra:cfn23sba}.
Unfortunately, there are some omissions and redundancies in that
paper, probably because the authors considered only the
indecomposable \sa s given in \cite{backh:class} and in some cases
did not take into account all isomorphisms to reduce the list of
\mst s.

In this paper we give lists of nonisomorphic real \mst s in
dimensions four and six, and classify the corresponding \dsd s. We
consider only the cases with nontrivial odd and even parts as purely
even \dd s were classified in
\cite{gom:ctd},\cite{hlasno:pltdm2dt},\cite{snohla:ddoubles} and
classification of the purely odd ones is trivial as the
superalgebras are (super)abelian.
\section{Lie superalgebras, \mst s and \dsd s}
The real Lie superalgebra $\cS$ is defined\cite{ro:supman} as real
$Z_2$--graded vector space $V=V_0\dot + V_1$ provided with Lie
superbracket $[\ ,\ ]$ satisfying
\begin{equation}\label{axsa1}
    [x,y]=-(-1)^{|x||y|}[y,x],
\end{equation}
\begin{equation}\label{axsa2}
    (-1)^{|x||z|}[x,[y,z]]+(-1)^{|y||x|}[y,[z,x]]+(-1)^{|z||y|}[z,[x,y]]=0,
\end{equation}
where $x,y,z\in V_0\cup V_1$ are so called homogeneous elements of
$V$ and
\begin{equation}\label{grad}   |x|:=0\ {\rm if}\ x\in V_0,\ |x|:=1\ {\rm if}\ x\in V_1.\
\end{equation}
We say that the \sa {} is of the {\em superdimension} $(m,n)$ iff
$dimV_0=m$ and $dimV_1=n$. We can
always choose so called homogeneous basis $\{X_I\}$ in $V$

\[ \{X_I\}_{I=1}^{m+n}=\{ b_i, f_\alpha\}_{i,\alpha=1}^{m,n},\ \ |b_i|=0,|f_\alpha|=1. \]

The indecomposable superalgebras up to the dimension 4 were classified in \cite{backh:class}. In
the Tables 
\ref{tab01},\ref{tab20},\ref{tab10} we list all two and
three--dimensional superalgebras with nontrivial odd
part\footnote{By $A_{m,n}$ we denote the Abelian superalgebra of the
superdimension $(m,n)$.}.
 All of them are solvable. For later use we include also the decomposable ones as they can produce
indecomposable \mst s.

A bilinear form $\langle .,.\rangle$
 on $\cS$ is called {\em supersymmetric} iff $$\langle
x,y\rangle=(-1)^{|x||y|}\langle y,x\rangle
$$
and it is called {\em super ad-invariant} iff
\begin{equation}\label{adinvance}
    \langle [x,y],z\rangle+(-1)^{|x||y|}\langle y,[x,z]\rangle=0.
\end{equation}

To evade the problems with definitions of supergroups we shall
define the \dsd s only on the algebraic level. The Lie \sa {} $\cD$
provided with bilinear nondegenerate supersymmetric  and super
ad-invariant form $\langle .,.\rangle$ will be called {\em \dsd} iff
it can be decomposed into a pair of maximally isotropic subalgebras
$\cS,\wt\cS$ such that $\cD=\cS\dot+\wt\cS$. The triple
$(\cD,\cS,\wt\cS)$ is called {\em \mst}\cite{andru:lspls}. It
follows immediately from the properties of $\langle .,.\rangle$ that
$dim\ \cS =dim\ \wt\cS$ so that the dimension of $\cD$ is always
even.

Let $\cD$ and $\cD'$ are \dsd s with bilinear forms $\langle .,.\rangle$ and $\langle
.,.\rangle'$. They will be called isomorphic iff there is an isomorphism
$P:\cD\rightarrow\cD'$ of the \sa s such that
\[ (\forall x,y\in \cD)\ (\langle x,y\rangle=\langle Px,Py\rangle').\]
The \mst s $(\cD,\cS,\wt\cS)$, $(\cD',\cS',\wt\cS')$ will be called isomorphic iff there is
an isomorphism $P$ of their \dsd s such that
\[ P(\cS)=\cS',\ P(\wt\cS)=\wt\cS'. \]
Below we shall denote the \mst s $(\cD,\cS,\wt\cS)$ as $(\cS|\wt\cS)$.

In the following we shall assume that the \mst s  are so called {\em
boson -- fermion orthogonal} \footnote{The other, rather exotic,
possibilities were discussed in the bachelor thesis of J. Vysok\'y},
i.e.
\begin{equation}\label{bforth}
\langle \cS_0,\wt \cS_1\rangle=\langle \cS_1,\wt \cS_0\rangle=0.
\end{equation} It follows from the boson -- fermion orthogonality that the superdimensions of
the \sa s $\cS$ and $\wt\cS$ coincide, i.e. $(m,n)=(\tilde m,\tilde n)$, and in such \mst s we can choose {\em dual
homogeneous basis}
\begin{equation} \label{dhb}\{X_I,\wt X^J\}_{I,J=1}^{m+n}=
\{b_i,f_\alpha,\wt b^j,\wt
f^\beta\}_{i,\alpha,j,\beta=1}^{m,n,m,n}\end{equation} where  \[
    |b_i|=|\wt b^j|=0,\ |f_\alpha|= |\wt f^\beta|=1,
\]
\begin{equation}\label{dualbasis}
    \langle b_i,\wt b^j\rangle =\langle \wt b^j,b_i\rangle = \delta_i^j,\  \langle f_\alpha,\wt
    f^\beta\rangle=-\langle\wt
    f^\beta, f_\alpha\rangle=\delta_\alpha^\beta
\end{equation}
\[ \langle b_i,b_j\rangle =  \langle b_i, f_\alpha\rangle =  \langle b_i, \wt f^\beta\rangle =
 \langle f_\alpha,b_j\rangle =  \langle f_\alpha,f_\beta\rangle =  \langle f_\alpha, \wt b^j\rangle=0\\
\]\[ \langle \wt b^j,\wt b^i\rangle =  \langle \wt b^j,f_\alpha\rangle =  \langle \wt b^j,\wt
f^\beta\rangle= \langle \wt f^\beta,b_j\rangle =  \langle \wt f^\beta, \wt b^j\rangle =
\langle \wt f^\beta, \wt f^\alpha\rangle =0 \] The block matrix of the bilinear form
$\langle.,.\rangle$ in this basis is
\begin{equation}\label{mbform}
    B=\left(
\begin{array}{cccccccc}
 0 & 0 & \unit_m & 0 \\
 0 & 0 & 0 & \unit_n \\
 \unit_m & 0 & 0 & 0 \\
 0 & -\unit_n & 0 & 0
\end{array}
\right),
\end{equation}
where $\unit_k$ is the identity matrix of dimension $k$. It is
obvious that the superdimension of these \mst s and \dsd s is
$(2m,2n)$.

A special type of the \dsd {} isomorphism is {\em T--duality} that
is the linear \tfn {} $T:\cD\rightarrow\cD$
\[T:b_i\mapsto \wt b^i,\,f_\alpha\mapsto\wt f^\alpha,\,\wt b^j\mapsto b_j,\,\wt f^\beta\mapsto -f_\beta.\]
Its \tfn {} \mtx {} is equal to $B$. Clearly
$T(\cS|\wt\cS)=(\wt\cS|\cS)$ and we shall call the \mst {}
$(\wt\cS|\cS)$ {\em dual to} $(\cS|\wt\cS)$. They are not isomorphic
in general.

Due to the super ad-invariance of $\langle.,.\rangle$ structure coefficients of the \mst {}
$(\cS|\wt\cS)$ in the dual basis are given by structure coefficients of the subalgebras $\cS$
 and $\wt\cS$
\begin{equation}\label{sc12}
    [X_I,X_j]={F_{IJ}}^K X_K, \ [\wt X^I, \wt X^J]=  {\wt F{^{IJ}}_K} \wt X^K
\end{equation}
as
\begin{equation}\label{mixedsc}
   [X_I,\wt X^J]= {\wt F{^{JK}}_I} X_K + {F_{KI}}^J \wt X^K.
\end{equation}
Jacobi identities (\ref{axsa2}) for the \dsd {} then imply the compatibility conditions for
the subalgebras of the \mst {} $(\cS|\wt\cS)$.

We are going to classify the \mst s and \dsd s with nontrivial odd
and even parts. As stated above their dimension is always even and
two dimensional cases are (super)abelian. It means that the simplest
interesting cases have dimensions four and six.
\section{Method of classification}
To classify the \mst s and \dsd s of superdimension $(2m,2n)$ we
start with a \sa {} $\cS$ of the superdimension $(m,n)$ and look for
structure coefficients of dual \sa {}  $\wt\cS$ by solving  Jacobi
identities
\begin{equation}\label{jac2}
    (-1)^{|\wt X^I||\wt X^K|} [\wt X^I,[\wt X^J,\wt X^K]] + cyclic\{I,J,K\}=0
\end{equation}
and
\begin{equation}\label{mixjac}
    (-1)^{| \wt X^I|| X_K|} [\wt X^I,[\wt X^J, X_K]] +
    cyclic\{I,J,K\}=0.
\end{equation}
We assume that the super Lie brackets are in the form
(\ref{sc12},\ref{mixedsc}) that means that we work in the dual
homogeneous bases (\ref{dhb}) of \mst s.

Next step is search for a list of different classes of isomorphic
\mst s and a choice of "canonical" forms of structure coefficients
${\wt F{^{IJ}}_K}$ for each class. The isomorphic triples are
related by the \tfn s of dual homogeneous bases that can be written
in the block form as
\begin{equation}\label{isommst}
    \left(
\begin{array}{c}
 X' \\
\wt X'
\end{array}
\right)=
    \left(
\begin{array}{cc}
 A & 0 \\
0 & (A^{-1})^T
\end{array}
\right)\left(
\begin{array}{c}
 X \\
\wt X
\end{array}
\right),
\end{equation}
where $(X,\wt X)^T$ and $(X',\wt X')^T$  are dual homogeneous bases
of the \mst s and $A$'s are (block diagonal) matrices of the group
of automorphisms of the \sa {} $\cS$.

Finally we classify the \dsd s by looking for the classes of \mst s that are isomorphic as
\dsd s. By definition the \dsd s $\cD$ and $\cD'$ are isomorphic if there is a linear
bijection that transforms grading, bilinear form and algebraic structure of $\cD$ into those
of $\cD'$. If dual homogeneous bases are chosen in both of them then it means that there is
$2(m+n)\times 2(m+n)$ \mtx {} $C$ of the block form
\begin{equation}\label{mtxC}
    C=\left(
\begin{array}{cc}
 P & Q \\
R & U
\end{array}
\right)\end{equation} where $P,Q,R,U$ are $(m+n)\times(m+n)$ block
diagonal matrices and elements of $C$ satisfies \eqn s
\begin{equation}\label{cpodm}
{C_a}^p {C_b}^q B_{pq}= B_{ab},\  {C_a}^p {C_b}^q {
\cF_{pq}}^r ={ \cF'_{ab}}^c {C_c}^r,
\end{equation} where $B$ is the \mtx {} (\ref{mbform}) and $\cF,\cF'$ are structure coefficients
of the \dsd s $\cD$ and $\cD'$. \section{Classification in
superdimensions $(2,2)$, $(4,2)$ and $(2,4)$} As mentioned in the
Introduction it is relevant to consider only the superalgebras with
nontrivial both even and odd part i.e. $dim V_0,\ dim V_1\geq 1$,
i.e. the \mst s and \dsd s of superdimensions $(2,2)$, $(4,2)$ and
$(2,4)$. Applying the method from the previous Section we get the
following results.
\subsection{\mst s and \dsd s of the superdimension $(2,2)$}
The maximal isotropic subalgebras $\cS$ and $\wt\cS$ of \mst s of
superdimension $(2,2)$ must have superdimensions $(1,1)$. They must
be isomorphic to those given in the Table \ref{tab01} and one can
choose the super Lie brackets of $\cS$ in the form given there.

\begin{table}[h]
\caption{Nonisomorphic Lie \sa s of the superdimension (1,1)}
{\begin{tabular}[c]{|c|c|c|c|l|} \hline
 Name & Nonzero super Lie brackets     & \\ \hline
 $ \Ajj $     &                      & Abelian    \\  & & \\
 $ \Njj=q(1) $     &  $ [f_1,f_1]= b_1 $  &  Nilpotent \\  & & \\
 $ \Sjj=P(1) $     &  $ [b_1,f_1]= f_1 $  &  Solvable  \\ & & \\
        \hline
\end{tabular}}
\label{tab01}
\end{table}

For each $\cS$ from the Table \ref{tab01} it is then rather easy to calculate the structure coefficients of the dual
algebras $\wt \cS$ satisfying (\ref{jac2}) and (\ref{mixjac}) and use the groups of automorphisms of $\cS$ for finding
five classes of nonisomorphic \mst s $(\cS|\wt\cS)$ of the superdimension $(2,2)$. They are listed in the Table
\ref{tab2}. Note that even though the \sa {} $\Ajj$ is decomposable the semiabelian \mst s $ (\Njj|\Ajj)
 $ and $ (\Sjj|\Ajj) $ are indecomposable.
Traditional notation for $(C^2_{-1}+A)$ is $gl(1|1)$ and the T--dual
to the \mst {} 4 (i.e. $\epsilon =+1$ appears in \cite{olshan} as a
bialgebra on $q(1)$.

\begin{table}[h]
\caption{List of nonisomorphic \mst s of the superdimension $(2,2)$ up to T--duality} {\begin{tabular}[c]{|c|c|c|c|l|}
\hline & & & \\ &$(\cS|\wt\cS)$& Nonzero super Lie brackets     & Backhouse's \cfn \\ & & & \\
\hline  & & & \\
   1& $ (\Ajj|\Ajj) $     &                      & Abelian    \\
   & & & \\

   2& $ (\Njj|\Ajj) $     &  $ [f_1,f_1]= b_1 $, $ [f_1,\wt{b^1}]= \wt{f^1} $, &  $(\ct+A)$   \\
         & & & \\
   3& $ (\Sjj|\Ajj) $     &  $ [b_1,f_1]= f_1 $, $ [b_1,\wt{f^1}]= -\wt{f^1} $, & $(C^2_{-1}+A)$   \\
         & & $ [f_1,\wt{f^1}]= \wt{b^1}$& \\
         & & & \\

   4,5& $ (\Sjj|\Njjep) $     &  $ [b_1,f_1]= f_1 $, $[b_1,\wt{f^1}]= \epsilon\, f_1-\tfj $,  &    $\cong(C^2_{-1}+A)$\\
        & $\epsilon=\pm 1$&  $ [f_1,\wt{f^1}]= \tbj $, $ [\wt{f^1},\wt{f^1}]= \epsilon\, \tbj $ &\\
        & & & \\
\hline
\end{tabular}}
\label{tab2}
\end{table}

As indicated in the Table \ref{tab2}, the \mst s $ (\Sjj|\Ajj) $ and
$(\Sjj|\Njjep)$ are isomorphic as \sa s and one can find an
isomorphism that preserve also the bilinear form $\langle
.,.\rangle$ so that they belong (together with their T--duals)
 to the same \dsd.
The isomorphism of \mst s between $ (\Sjj|\Ajj) $ and
$(\Sjj|\Njjep)$ is given by the matrix
\[ C= \left(
\begin{array}{cccc}
 1& 0& 0& 0 \\
 0& 1& 0& 0 \\
 0& 0& 1& 0 \\
 0& \epsilon/2& 0&  1
\end{array} \right),
\]i.e. by the \tfn {}
\begin{equation}\label{tfn11}
    (b_1',f_1',\tbj',\tfj')=(b_1,f_1,\tbj,\tfj+\half\epsilon f_1),
\end{equation} where $(b_1,f_1,\tbj,\tfj)$ are generators of the \mst {} $ (\Sjj|\Ajj) $
and $(b_1',f_1',\tbj',\tfj')$  are generators of the \mst s $
(\Sjj|\Njjep) $. We get the folowing theorem.
\begin{theorem}  \label{veta11} Any real Drinfeld
superdouble of the superdimension (2,2) belongs just to one of the
following 3 classes and allows decomposition into all \mst s listed
in the class and their duals $(\cS \leftrightarrow \wt\cS)$.
\begin{description}
    \item[$DD_{(2,2)}\ I$]  :  $ (\Ajj|\Ajj) $,
    \item[$DD_{(2,2)}\ II$] : $ (\Njj|\Ajj) $,
    \item [$DD_{(2,2)}\ III$] : $ (\Sjj|\Ajj) $, $ (\Sjj|\Njjep),\ \epsilon=\pm 1
    $.
\end{description}\end{theorem}
\subsection{\mst s and \dsd s of the superdimension $(4,2)$}
The real Lie \sa s of the superdimension (2,1) and their groups of
automorphisms are listed\footnote{Differently from
\cite{backh:class} we have denoted the superalgebra $C^1_\half$ as
$F$ because it differs from $C^1_{p=\half}$.} in the Table
\ref{tab20}.
\begin{table}[h]
\caption{Three--dimensional superalgebras of the superdimension (2,1)}
\begin{tabular}[c]{|c|c|c|c|l|} \hline
Name & Nonzero super Lie brackets     &  Group of automorphisms & Comments \\
\hline\hline $ \Adj $     &                      &
$\left(\begin{array}{ccc}
a&b&0  \\ c & d &0  \\ 0&0&k\end{array} \right) $ & \\

 & & & \\

$ \Ndj $     &  $ [f_1,f_1]= b_1 $  & $\left(\begin{array}{ccc}
d^2&0&0  \\ b & a &0 \\ 0&0&d\end{array} \right) $ & $=\Njj\oplus\Ajn$\\

 & & & \\

$ \Sdj $     &  $ [b_1,f_1]= f_1 $  & $\left(\begin{array}{ccc}
1&b&0  \\ 0 & c &0  \\ 0&0&d\end{array} \right) $ & $=\Sjj\oplus\Ajn$\\

 & & & \\

$ \cjp $     &  $ [b_1,b_2]= b_2 $, $ [b_1,f_1]= p\,f_1$ &
$\left(\begin{array}{ccc}
1&b&0  \\ 0 & c &0  \\ 0&0&d\end{array} \right) $ & $p\in \real$\\

 & & & \\

$ \chalf $     &  $ [b_1,b_2]= b_2 $, $ [b_1,f_1]= \half\,f_1 $,  $
[f_1,f_1]= b_2$ & $\left(\begin{array}{ccc}
1&b&0 \\ 0 & d^2 &0  \\ 0&0&d\end{array} \right) $ & \\
\hline
\end{tabular}
\label{tab20}
\end{table}

By checking the Jacobi identities and employing the groups of
automorphisms of the \sa {} $\cS$ we find that there are 14 classes
of real \mst s of the superdimension (4,2) up to the T--duality.
They are listed in the Table \ref{tab21b}.

\begin{table}[h]
\caption{List of nonisomorphic \mst s of the superdimension $(4,2)$ up to T--duality.}{\begin{tabular}[c]{|c|c|c|c|c|}
\hline  & & &  & \\ $MT_{(4,2)}$&  $\cS$&$\wt\cS$ & Nonzero super Lie brackets & Comments \\
\hline\hline
& & & & \\

1& $\Adj $ & $\Adj$ & &  \\
& & & & \\

\hline

& & & & \\
& $\Ndj $ & &$ [f_1,f_1]= b_1 $ &  \\
& & & & \\
\hline
& & & & \\

2 & & $\Adj$ & & \\
& & & & \\

\hline

& & & & \\
& $\Sdj $ & & $ [b_1,f_1]= f_1 $ &  \\
& & & & \\
\hline
& & & & \\

3 & & $\Adj$ & & \\
& & & & \\

4 & & $\Ndjep$ & $ [\tfj,\tfj]= \epsilon\,\tbj $ & $\epsilon=\pm 1$ \\
& & & & \\

5 & & $\Sdj$ & $ [\tb2,\tfj]= \tfj $ & \\
& & & & \\

\hline

& & & & \\
& $\cjp $ & & $ [b_1,b_2]= b_2 $, $ [b_1,f_1]= p\,f_1 $ & $p\in\real$ \\
& & & & \\
\hline
& & & & \\
6 & & $\Adj$ & & \\
& & & & \\

7 & & $\Ndjep$ & $ [\tfj,\tfj]= \epsilon\,\tbj $ & $\epsilon=\pm 1$ \\
& & & & \\

8 & & $\wt C^1_{-p}$ & $[\tbj,\tb2]= \tbj$, $[\tb2,\tfj]= p\,\tfj$& \\
& & & & \\
\hline
& & & & \\
9 & & $\wt\Ndj$ & $ [\tfj,\tfj]= \tb2$ & $ p=\half$ \\
& & & & \\

10 & & $\cjn,\kappa$ & $ [\tbj,\tb2]= \kappa\,\tb2$ & $ p=0,\ \kappa\neq 0$ \\
& & & & \\

\hline
& & & & \\
& $\chalf $ & & $ [b_1,b_2]= b_2 $, $ [b_1,f_1]= \half\,f_1 $, $[f_1,f_1]=b_2 $ & \\
& & & & \\
\hline
& & & & \\
11 & & $\Adj$ & & \\
& & & & \\

12 & & $\cjhalfep$ & $ [\tbj,\tb2]= \epsilon\,\tbj$, $ [\tb2,\tfj]= \half\epsilon\,\tfj$ & $\epsilon=\pm 1$ \\
& & & & \\

13 & & $\chalf.i,\epsilon$ & $ [\tbj,\tb2]= \epsilon\,\tbj$, $
[\tb2,\tfj]= -\half \epsilon\,\tfj$, $
[\tfj,\tfj]=\epsilon\,\tbj$ & $\epsilon=\pm 1$ \\
& & & & \\

14 & & $\chalf. ii,\kappa$ & $ [\tbj,\tb2]= \kappa\,\tb2$, $
[\tbj,\tfj]= \half \kappa\,\tfj$,
$ [\tfj,\tfj]= \kappa\,\tb2$ &  $\kappa\neq 0$\\
& & & & \\

\hline
\end{tabular}}
 \label{tab21b}
\end{table}

To classify the \dsd s it is useful to determine the dimensions of
multiple commutants $\cmt_1=[\cD,\cD],\ \cmt_2=[\cmt_1,\cmt_1],\
\ldots$ of the \mst s. This sorts them out into eight classes that
finally turn out to be the \dsd s. The only exception are the \mst s
with $dim\,\cmt_1=3,\ dim\,\cmt_2=1$ but the superdimensions of
their subalgebras $\cmt_1$ are (1,2) and (3,0) so that they belong
to nonisomorphic \dsd s. The classes are displayed in the Table
\ref{tab25}. The \dsd {} isomorfisms among the corresponding \mst s
are given in the Appendix A.

\begin{table}[h]
\caption{Invariants of the \mst s of the superdimension (4,2) }{\begin{tabular}[c]{|c|c|c|c|l|} \hline
Dim. of     &  \mst s &  \dsd \\
${\cmt}_1,{\cmt}_2,{\cmt}_3$ & $MT_{(4,2)}$ & $DD_{(4,2)}$\\ \hline
\hline
0,\ 0,\ 0 & $(\Adj|\Adj)$ & $I$ \\ & & \\
2,\ 0,\ 0 & $(\Ndj|\Adj)$ & $II$ \\ & & \\
3,\ 1,\ 0 & $(\Sdj|\Adj)$, $(\Sdj|\Ndjep)$ , $(\Sdj|\Sdj)$ & $III$ \\ & & \\
3,\ 1,\ 0 & $(\cjn|\Adj)=(\cjn|\cjn,\kappa)$, $(\cjp|C^1_{-p})$, $p=\kappa=0$ & $IV_0$ \\ & & \\
3,\ 3,\ 3 & $(\cjn|\cjn,\kappa)$ & $IV_\kappa,\ \kappa\neq 0$\\ & & \\
4,\ 1,\ 0 & $(\cjn|\Ndjep)$& $V$ \\ & & \\
5,\ 1,\ 0 & $(\cjp|\Adj)$, $ (\cjp|\Ndjep) $, $(\cjp|C^1_{-p})$ & $VI_p, \ p >0$ \\ & & \\
5,\ 3,\ 0 & $(\chalf|\Adj)$, $(C^1_{p=\half}|\Ndj)$, $(\chalf|\chalf.i,\epsilon)$, $(\chalf|\cjhalfep)$ & $VII$\\ & & \\
5,\ 5,\ 5 & $(\chalf|\chalf. ii,\kappa)$ & $VIII_\kappa$
\\ & & \\
\hline
\end{tabular}}
\label{tab25}\end{table}

\begin{theorem}  \label{veta21} Any real Drinfel'd
superdouble of the superdimension (4,2) belongs just to one of the
following classes and allows decomposition into nonisomorphic \mst s
listed in the class and their duals $(\cS \leftrightarrow \wt\cS)$.
\begin{description}
    \item[$DD_{(4,2)}\ I$]  :  $ (\Adj|\Adj) $,
    \item[$DD_{(4,2)}\ II$] : $ (\Ndj|\Adj) $,
    \item [$DD_{(4,2)}\ III$] : $ (\Sdj|\Adj) $, $ (\Sdj|\Ndjep) $, $ (\Sdj|\Sdj) $,
    \item[$DD_{(4,2)}\ IV_0$]  :  $(\cjn|\Adj)=(\cjn|\cjn,\kappa)$, $(\cjp|C^1_{-p})$, $p=\kappa= 0$,
    \item[$DD_{(4,2)}\ IV_\kappa$]  :  $(\cjn|\cjn,\kappa)$, $\kappa\neq 0$,
    \item[$DD_{(4,2)}\ V$]  :   $(\cjn|\Ndjep)$,
    \item[$DD_{(4,2)}\ VI_p$] : $(C^1_{\pm p}|\Adj)$, $ (C^1_{\pm p}|\Ndjep) $, $(\cjp|C^1_{-p})$, $p > 0$,
    \item [$DD_{(4,2)}\ VII$] : $(\chalf|\Adj)$, $(C^1_{p=\half}|\wt\Ndj)$, $(\chalf|\chalf.i,\epsilon)$, $(\chalf|\cjhalfep)$,
    \item [$DD_{(4,2)}\ VIII_\kappa$] : $(\chalf|\chalf. ii,\kappa)$.
\end{description}\end{theorem}
\subsection{\mst s and \dsd s of the superdimension $(2,4)$}
The real Lie \sa s of the superdimension (1,2) togeher with their
groups of automorphisms are listed in the Table \ref{tab10}.
\begin{table}[h]
\caption{Superalgebras of the  superdimension (1,2) } {\begin{tabular}[c]{|c|c|c|c|l|}
\hline Name & Nonzero super Lie brackets     & Group of automorphisms \\
\hline\hline  $ \Ajd $     &                      &
$\left(\begin{array}{ccc}
k&0& 0 \\ 0&a&b \\ 0 & c & d \end{array} \right) $ \\

 & & \\

 $N_{12}^0$ &  $ [f_1,f_1]= b_1 $& $\left(\begin{array}{ccc}
a^2&0& 0 \\ 0&a & b \\ 0 & 0 & d \end{array} \right) $  \\

 & & \\

 $\begin{array}{c} \Njdep \\ \epsilon=\pm 1\end{array}$ &  $ [f_1,f_1]= b_1 $, $ [f_2,f_2]= \epsilon\,b_1 $ & $\left(\begin{array}{ccc}
d^2+\epsilon\, c^2&0& 0 \\ 0&\mp\epsilon\, d & \pm c \\ 0 & c & d \end{array} \right) $  \\

 & & \\

$ C^2_{-1}$     &  $ [b_1,f_1]= f_1 $, $ [b_1,f_2]= - f_2 $ &
$\left(\begin{array}{ccc} -1&0& 0 \\ 0&0&b \\ 0 & c & 0 \end{array}
\right), \left(\begin{array}{ccc}
1&0& 0 \\ 0&c& 0 \\ 0 & 0 & d \end{array} \right) $   \\

 & & \\

$ \begin{array}{c}\cdp\\ -1<p<1 \end{array}$     &  $ [b_1,f_1]= f_1
$, $ [b_1,f_2]= p\, f_2 $ & $\left(\begin{array}{ccc}
1&0& 0 \\ 0&c& 0 \\ 0 & 0 & d \end{array} \right) $   \\

 & & \\

$ C^2_{1}$     &  $ [b_1,f_1]= f_1 $, $ [b_1,f_2]=  f_2 $ &
$\left(\begin{array}{ccc}
1&0& 0 \\ 0&a&b \\ 0 & c & d \end{array} \right) $   \\

 & & \\

$ \ct $     &  $ [b_1,f_2]= f_1 $  & $\left(\begin{array}{ccc}
a&0& 0 \\ 0& a d&0 \\ 0 & c & d \end{array} \right) $   \\

 & & \\

 $ \cc $     &  $ [b_1,f_1]= f_1 $, $ [b_1,f_2]= f_1+f_2 $   &  $\left(\begin{array}{ccc}
1&0& 0 \\ 0&a&0 \\ 0 & c & a \end{array} \right) $  \\

 & & \\

$ \begin{array}{c}\cpp\\ p> 0 \end{array}$    &  $ [b_1,f_1]=
p\,f_1-f_2 $, $ [b_1,f_2]= f_1+p\,f_2 $ & $\left(\begin{array}{ccc}
 1&0& 0 \\ 0&a& -c \\ 0 & c & a \end{array} \right)$   \\
 & & \\

$\cpn$ & $ [b_1,f_1]=-f_2 $, $ [b_1,f_2]= f_1$ & $\left(\begin{array}{ccc} \pm 1&0& 0 \\
0&\pm a&\mp c \\ 0 & c & a \end{array} \right) $   \\

\hline
\end{tabular}}
\label{tab10}
\end{table}

The Jacobi identities imply that the \mst s of the superdimension
$(2,4) $ are only of the forms $(C|\Nabg)$ or their T--duals, where
$C=\Ajd,\cdp,\ct,\cc,\cpp$, and $\Nabg$ are \sa s with super Lie
products
\begin{equation}\label{nabg}
[\tfj,\tfj]= \alpha\,\tbj,\ [\tfj,\tf2]= \beta\,\tbj,\ [\tf2,\tf2]=
\gamma\,\tbj.
\end{equation} They are isomorphic to $\Njdep=N_{12}^{1,0,\epsilon}$ or $N^0_{12}=N_{12}^{1,0,0}$ or
 $\Ajd=N_{12}^{0,0,0}$.

Using the automorphisms  of $\Ajd,\cdp,\ct,\cc,\cpp$ displayed in
the Table \ref{tab10} we find that there are 31 classes of \mst s of
the superdimension (2,4) up to the T--duality. They  are listed in
the Table \ref{tab12}.
\begin{table}[h]
\caption{List of nonisomorphic \mst s of the superdimension $(2,4)$ up to T--duality,
 $\epsilon,\epsilon_1,\epsilon_2=\pm 1,\ \delta=0,1.$}  {\begin{tabular}[c]{|c|c|c|c|c|}
\hline  & & &  \\ $MT_{(2,4)}$ & $\cS$&$\wt\cS $  & Comments \\
\hline\hline
 & & &  \\

1--3 & $\Ajd $ &  $\Ajd,\ N_{12}^{1,0,0},\ N_{12}^{1,0,\epsilon}$ &
\\
 & & & \\

4 -- 8& $\cdp $ &  $\Ajd,\ N_{12}^{0,1,0},\
N_{12}^{0,\delta,\epsilon},\ N_{12}^{\epsilon,\delta,0}, \
N_{12}^{\epsilon_1,\kappa,\epsilon_2}$ & $|p|<1,\ \kappa\geq 0$
\\
 & & & \\

9 -- 12& $C^2_1  $ &  $ \Ajd,\ N_{12}^{\epsilon,0,\epsilon},\
N_{12}^{1,0,-1}, \ N_{12}^{0,0,\epsilon} $ &
\\
 & & & \\

13 -- 17& $C^2_{-1} $ &  $\Ajd,\ N_{12}^{0,1,0},\
N_{12}^{0,\delta,\epsilon},\
 \ N_{12}^{1,\kappa,1},\ N_{12}^{\epsilon,\kappa,-\epsilon} $ & $\kappa\geq 0$
\\
 & & & \\
18 -- 22& $\ct$ &  $\Ajd,\ N_{12}^{\epsilon,0,1},\ N_{12}^{1,0,0},\ N_{12}^{0,\epsilon,0} ,\ N_{12}^{0,0,1} $ &
\\
 & & & \\
23 -- 26 & $\cc$ &  $\Ajd,\ N_{12}^{\epsilon,0,0},\ N_{12}^{0,\epsilon,0},\ N_{12}^{\kappa,0,\epsilon} $ &
$\kappa\in\real$
\\
 & & & \\
27 -- 29 & $\cpp$ &  $\Ajd,\ N_{12}^{\kappa,0,\epsilon},\ N_{12}^{-1,0,-1} $ & $p>0,\ -1<\kappa\leq 1$
\\
 & & & \\
30, 31 & $C^5_{p=0}$ &  $\Ajd,\  N_{12}^{\kappa,0,1} $ & $-1\leq\kappa\leq 1$
\\
 & & & \\
\hline
\end{tabular}}
\label{tab12}
\end{table}

We shall show that in many cases the \mst s $(C|\Nabg)$ belong to
the same \dsd {} as the semiabelian $(C|\Ajd)$, in other words,
there are linear transformations between the basis
$(b_1,f_1,f_2,\tbj,\tfj,\tf2)$ of $(C|\Ajd)$ and $
(b_1',f_1',f_2',\tbj',\tfj',\tf2')$ of $(C|\Nabg)$ that preserve the
bilinear form $\langle.,.\rangle$ and transform the super Lie
brackets of $(C|\Ajd)$ into those of $(C|\Nabg)$.

Let the \mst {} $(C|\Nabg)$ is given by  super Lie brackets of $\cS$
and $\wt\cS$
\begin{equation}\label{cnjd}
    [b_1',f_j']= {H_j}^k f_k' , \ \ [{\wt{f^j}}',{\wt{f^k}}']=G^{jk}\tbj',\ j,k\in\{1,2\},
\end{equation}
i.e. \[G^{jk}=\left(
\begin{array}{cc}
 \alpha& \beta \\
 \beta&\gamma
\end{array}
\right).\] The only nonzero super Lie brackets of $(C|\Ajd)$ are
\begin{equation}\label{cajd}
    [b_1,f_j]= {H_j}^k f_k ,\ \ [f_j,{\wt{f^k}}]={H_j}^k\tbj,\ \ [b_1,{\wt{f^k}}]=-{H_j}^k\wt{f^j}.
\end{equation}
Let the transformation of bases is given by the symmetric matrix
$R^{jk}$
\begin{equation}\label{tfprime}
(b_1',f_1',f_2',\tbj',\tfj',\tf2')=(b_1,f_1,f_2,\tbj,\tfj+R^{1k}f_k,\tf2+R^{2k}f_k).
\end{equation}
It is easy to see that this \tfn {} preserves the bilinear form
$\langle.,.\rangle$ and
\begin{equation}\label{tfliep}
    [{\wt{f^j}}',{\wt{f^k}}']=(R^{jl}{H_l}^k + R^{kl}{H_l}^j )\tbj'.
\end{equation}
The super Lie brackets of (\ref{cnjd}) then can be obtained by
solving $R^{jk}$ from the \eqn
\begin{equation}\label{eqforR}
   R^{jl}{H_l}^k + R^{kl}{H_l}^j =G^{jk}.
\end{equation}
It can be solved for general $G^{jk}$ iff $Trace\ H \neq 0$ and $
det\ H \neq 0$. These conditions are satisfied for $\cc$, $\cdp,\
p\neq -1,0$ and $\cpp,\ p \neq 0$. In the exceptional cases the \eqn
{} (\ref{eqforR}) can be solved for particular values of $G^{jk}$,
namely, for $\cdn$ and $\ct$ only if $G^{33}=\gamma=0$, for
$C^2_{-1}$ only if $G^{23}=\beta=0$, and for $\cpn$ only if
$G^{33}=\gamma=-G^{22}=-\alpha$. These cases are not isomorphic to
the semiabelian \mst s $(C|\Ajd)$, but e.g. to $(C|N_{12}^{0,0,1})$.
The isomorphisms are given in the Appendix B.

We get the folowing theorem.
\begin{theorem}  \label{veta12} Any real Drinfel'd
superdouble of the superdimension (2,4) belongs just to one of the
following classes and allows decomposition into all \mst s listed in
the class and their duals $(\cS \leftrightarrow \wt\cS)$.
\begin{description}
    \item[$DD_{(2,4)}\ I$]  :  $ (\Ajd|\Ajd) $,

    \item[$DD_{(2,4)}\ II_p$]  :  
    $ (C^2_{\pm p}|\Nabg) $,
    $0<p<1$,

    \item[$DD_{(2,4)}\ II_1$]  :  
    $ (C^2_1|\Nabg) $, 
    $ (C^2_{-1}|N_{12}^{\lambda,0,\kappa}) $,

    \item[$DD_{(2,4)}\ II_0$]  :  
    $ (C^2_0|N_{12}^{\alpha,\beta,0}) $,

     \item[$DD_{(2,4)}\ III $]  :  
     $ (\ct|N_{12}^{\alpha,\beta,0}) $, $ (\Ajd|N^{\lambda,\kappa,\gamma}_{12}),\ \lambda\gamma<\kappa^2 $,

    \item[$DD_{(2,4)}\ IV $]  :  
    $ (\cc|\Nabg) $,
    $ (C^2_{-1}|N_{12}^{\lambda,\kappa,\rho}) $, $\kappa\neq 0$,

    \item[$DD_{(2,4)}\ V_{p}$]  :  
    $(\cpp|N_{12}^{\alpha,\beta,\gamma})$, $p>0$,

    \item[$DD_{(2,4)}\ V_0 $]  :  
    $ (C^5_0|N_{12}^{\alpha,\beta,-\alpha})$,

    \item[$DD_{(2,4)}\ VI $]  :  
    $ (C^2_0|N_{12}^{\alpha,\beta,\gamma})  $, $\gamma\neq 0$,

    \item[$DD_{(2,4)}\ VII $]  :  
$(\ct|N_{12}^{\alpha\beta,\gamma})$, $\gamma\neq 0$,

    \item[$DD_{(2,4)}\ VIII $]  :  
    $ (C^5_0|N_{12}^{\alpha,\beta,\gamma}) $,
    $\alpha\neq -\gamma$,

   \item[$DD_{(2,4)}\ IX$]  :   $ (\Ajd|N^{\alpha,\beta,\gamma}_{12}),\ \alpha\gamma>\beta^2
   $,

    \item[$DD_{(2,4)}\ X$]  :  $ (\Ajd|N^{\alpha,\beta,\gamma}_{12}),\
    ({\alpha,\beta,\gamma})\neq (0,0,0),\ \alpha\gamma=\beta^2 $.
   \end{description}\end{theorem}

Note that the \dsd s do not depend on the parameters of \mst s $\alpha,\beta,\gamma,\kappa\in \real$. The \dsd s
$II_p$ - $V_0$ and $IX,X$can be called semiabelian as they are isomorphic to $(\cS|\Ajd)$. Dimensions of commutants of
some \dsd s given above,
are equal and  we are able to prove their nonisomorphicity only by
the computer calculations.
\section{Conclusion}We have classified
real \mst s and \dsd s of the superdimensions (2,2), (4,2) and
(2,4), i.e real Lie \sa s provided with with bilinear nondegenerate supersymmetric and super ad-invariant form such
that they can be decomposed into a pair of maximally isotropic subalgebras.

It turned out that nearly all investigated \dsd s contain more than
one pair of nonisomorphic dual \mst s. This gives us the possibility
to apply the \pltp y to \sm s whose targets are supergroups.
\section*{Acknowledgement}
This work was supported by the research plan LC527 of the Ministry
of Education of the Czech Republic.
\section*{Appendix A. Isomorfisms of the \dsd s of the superdimension
(4,2)}\label{appA} ${\mathbf{ DD_{(4,2)} III}}$ :
\[ (\Sdj|\Adj)  \rightarrow (\Sdj|\Ndjep)\ :\ \ C=\left(
\begin{array}{cccccc}
 1 & 0 & 0 & 0 & 0 & 0 \\
 0 & 1 & 0 & 0 & 0 & 0 \\
 0 & 0 & 1 & 0 & 0 & 0 \\
 0 & 0 & 0 & 1 & 0 & 0 \\
 0 & 0 & 0 & 0 & 1 & 0 \\
 0 & 0 & \frac{\epsilon}{2} & 0 & 0 & 1
\end{array}
\right)
\]
\[ (\Sdj|\Adj)  \rightarrow (\Sdj|\Sdj)\ :\ \ C=\left(
\begin{array}{cccccc}
 1 & 0 & 0 & 0 & 0 & 0 \\
 0 & 1 & 0 & 0 & 0 & 0 \\
 0 & 0 & 1 & 0 & 0 & 0 \\
 0 & 1 & 0 & 1 & 0 & 0 \\
 -1 & 0 & 0 & 0 & 1 & 0 \\
 0 & 0 & 0 & 0 & 0 & 1
\end{array}
\right)
\]

${\mathbf{ DD_{(4,2)} IV_0}}$ :
\[ (\cjn|\Adj) \rightarrow (\cjp|C^1_{-p}),\ p=0 :\ \ C=\left(
\begin{array}{cccccc}
 1 & 0 & 0 & 0 & 0 & 0 \\
 0 & 1 & 0 & 0 & 0 & 0 \\
 0 & 0 & 1 & 0 & 0 & 0 \\
 0 & 1 & 0 & 1 & 0 & 0 \\
 -1 & 0 & 0 & 0 & 1 & 0 \\
 0 & 0 & 0 & 0 & 0 & 1
\end{array}
\right)
\]

${\mathbf{ DD_{(4,2)} V}}$ :
\[ (\cjn|N^+_{21}) \rightarrow (\cjn|N^-_{21})  :\ \ C=\left(
\begin{array}{cccccc}
-1 & 0 & 0 & 0 & 0 & 0 \\
 0 & 0 & 0 & 0 & 1 & 0 \\
 0 & 0 & 1 & 0 & 0 & 0 \\
 0 & 0 & 0 & -1 & 0 & 0 \\
 0 & 1 & 0 & 0 & 0 & 0 \\
 0 & 0 & 0 & 0 & 0 & 1
\end{array}
\right)
\]

${\mathbf{ DD_{(4,2)} VI_p}}$ :
\[ (\cjp|\Adj)  \rightarrow (C^1_{-p}|\Adj)\ :\ \ C=\left(
\begin{array}{cccccc}
 -1 & 0 & 0 & 0 & 0 & 0 \\
 0 & 0 & 0 & 0 & 1 & 0 \\
 0 & 0 & 1 & 0 & 0 & 0 \\
 0 & 0 & 0 & -1 & 0 & 0 \\
 0 & 1 & 0 & 0 & 0 & 0 \\
 0 & 0 & 0 & 0 & 0 & 1
\end{array}
\right)
\]
\[ (\cjp|\Adj)  \rightarrow (\cjp|\Ndjep)\ :\ \ C=\left(
\begin{array}{cccccc}
 1 & 0 & 0 & 0 & 0 & 0 \\
 0 & 1 & 0 & 0 & 0 & 0 \\
 0 & 0 & 1 & 0 & 0 & 0 \\
 0 & 0 & 0 & 1 & 0 & 0 \\
 0 & 0 & 0 & 0 & 1 & 0 \\
 0 & 0 & \frac{\epsilon}{2 p} & 0 & 0 & 1
\end{array}
\right)
\]
\[ (\cjp|\Adj)  \rightarrow (\cjp|C^1_{-p},\epsilon)\ :\ \ C=\left(
\begin{array}{cccccc}
 1 & 0 & 0 & 0 & 1 & 0 \\
 0 & 1 & 0 & -1 & 0 & 0 \\
 0 & 0 & 1 & 0 & 0 & 0 \\
 0 & 1 & 0 & 0 & 0 & 0 \\
 -1 & 0 & 0 & 0 & 0 & 0 \\
 0 & 0 & 0 & 0 & 0 & 1
\end{array}
\right)
\]
${\mathbf{ DD_{(4,2)} VII}}$ :
\[ (\chalf|\Adj)  \rightarrow (C^1_{p=\half}|\wt\Ndj)\ :\ \ C=\left(
\begin{array}{cccccc}
 -1 & 0 & 0 & 0 & 0 & 0 \\
 0 & 0 & 0 & 0 & 1 & 0 \\
 0 & 0 & 0 & 0 & 0 & 1 \\
 0 & 0 & 0 & -1 & 0 & 0 \\
 0 & 1 & 0 & 0 & 0 & 0 \\
 0 & 0 & -1 & 0 & 0 & 0
\end{array}
\right)
\]
\[ (\chalf|\Adj)  \rightarrow (\chalf|\chalf.i,\epsilon) \ :\ \ C=\left(
\begin{array}{cccccc}
 1 & 0 & 0 & 0 & {\epsilon } & 0 \\
 0 & 1 & 0 & -\epsilon  & 0 & 0 \\
 0 & 0 & \epsilon  & 0 & 0 & -1 \\
 0 & \epsilon  & 0 & 0 & 0 & 0 \\
 -{\epsilon } & 0 & 0 & 0 & 0 & 0 \\
 0 & 0 & 1 & 0 & 0 & 0
\end{array}
\right)
\]
\[ (\chalf|\Adj)  \rightarrow (\chalf|\cjhalfep) \ :\ \ C=\left(
\begin{array}{cccccc}
 1 & 0 & 0 & 0 &{\epsilon } & 0 \\
 0 & 1 & 0 & -\epsilon  & 0 & 0 \\
 0 & 0 & \epsilon  & 0 & 0 & -1 \\
 0 & \epsilon  & 0 & 0 & 0 & 0 \\
 -{\epsilon } & 0 & 0 & 0 & 0 & 0 \\
 0 & 0 & 0 & 0 & 0 & {\epsilon }
\end{array}
\right)
\]

\section*{Appendix B. Isomorfisms of the \dsd s of the superdimension
(2,4)}\label{appB}

${\mathbf{ DD_{(2,4)} II_p, \ p\neq -1,0}}$ :\[
(\cdp|\Ajd)\rightarrow (\cdp|\Nabg)\ :\ \ C = \left(
\begin{array}{cccccc}
 1 & 0 & 0 & 0 & 0 & 0 \\
 0 & 1 & 0 & 0 & 0 & 0 \\
 0 & 0 & 1 & 0 & 0 & 0 \\
 0 & 0 & 0 & 1 & 0 & 0 \\
 0 & \frac{\alpha }{2} & \frac{\beta }{p+1} & 0 & 1 & 0 \\
 0 & \frac{\beta }{p+1} & \frac{\gamma }{2 p} & 0 & 0 & 1
\end{array}
\right)\]

\[ (\cdp|\Ajd)\rightarrow (C^2_{-p}|\Ajd)\ :\ \ C = \left(
\begin{array}{cccccc}
 1 & 0 & 0 & 0 & 0 & 0 \\
 0 & 1 & 0 & 0 & 0 & 0 \\
 0 & 0 & 0 & 0 & 0 & -1 \\
 0 & 0 & 0 & 1 & 0 & 0 \\
 0 & 0 & 0 & 0 & 1 & 0 \\
 0 & 0 & 1 & 0 & 0 & 0
\end{array}
\right)\] ${\mathbf{ DD_{(2,4)} II_1}}$ :\[ (\cdj|\Ajd)\rightarrow
(\cdj|\Nabg)\ :\ \ C = \left(
\begin{array}{cccccc}
 1 & 0 & 0 & 0 & 0 & 0 \\
 0 & 1 & 0 & 0 & 0 & 0 \\
 0 & 0 & 1 & 0 & 0 & 0 \\
 0 & 0 & 0 & 1 & 0 & 0 \\
 0 & \frac{\alpha }{2} & \frac{\beta }{2} & 0 & 1 & 0 \\
 0 & \frac{\beta }{2} & \frac{\gamma }{2 } & 0 & 0 & 1
\end{array}
\right)\]
\[ (C^2_{-1}|\Ajd)\rightarrow (\cdj|\Ajd)\ :\ \ C = \left(
\begin{array}{cccccc}
1 & 0 & 0 & 0 & 0 & 0 \\
 0 & 1 & 0 & 0 & 0 & 0 \\
 0 & 0 & 0 & 0 & 0 & 1 \\
 0 & 0 & 0 & 1 & 0 & 0 \\
 0 & 0 & 0 & 0 & 1 & 0 \\
 0 & 0 & -1 & 0 & 0 & 0
\end{array}
\right)\] ${\mathbf{ DD_{(2,4)} II_0}}$ :
\[ (\cdn|\Ajd) \rightarrow (\cdn|N_{12}^{\alpha,\beta,0})\ :\ \ C =  \left(
\begin{array}{cccccc}
 1 & 0 & 0 & 0 & 0 & 0 \\
 0 & 1 & 0 & 0 & 0 & 0 \\
 0 & 0 & 1 & 0 & 0 & 0 \\
 0 & 0 & 0 & 1 & 0 & 0 \\
 0 & \frac{\alpha }{2} & \beta  & 0 & 1 & 0 \\
 0 & \beta  & 0 & 0 & 0 & 1
\end{array}
\right)\] ${\mathbf{ DD_{(2,4)} III}}$ :\[ (\ct|\Ajd) \rightarrow
(\ct|N_{12}^{\alpha,\beta,0})\ :\ \ C =  \left(
\begin{array}{cccccc}
 1 & 0 & 0 & 0 & 0 & 0 \\
 0 & 1 & 0 & 0 & 0 & 0 \\
 0 & 0 & 1 & 0 & 0 & 0 \\
 0 & 0 & 0 & 1 & 0 & 0 \\
 0 & 0 & \frac{\alpha }{2} & 0 & 1 & 0 \\
 0 & \frac{\alpha }{2} & \beta  & 0 & 0 & 1
\end{array}
\right)\]

\[ (\ct|\Ajd) \rightarrow (\Ajd|N_{12}^{\lambda,\kappa,\gamma})\
:\ \ C =  \left(
\begin{array}{cccccc}
 \rho ^2 & 0 & 0 & 0 & 0 & 0 \\
 0 & \frac{1}{2} & 0 & 0 & 0 & \gamma  \\
 0 & \frac{\rho -\kappa }{2 \gamma } & 0 & 0 & 0 & -\kappa -\rho  \\
 0 & 0 & 0 & \frac{1}{\rho ^2} & 0 & 0 \\
 0 & 0 & \frac{\kappa -\rho }{2 \gamma  \rho } & 0 & \frac{\kappa +\rho }{\rho } & 0 \\
 0 & 0 & \frac{1}{2 \rho } & 0 & \frac{\gamma }{\rho } & 0
\end{array}
\right),\ \rho=\sqrt{{\kappa^2-\lambda\gamma}}\]

 ${\mathbf{ DD_{(2,4)} IV}}$ :
\[(\cc|\Ajd) \rightarrow (\cc|N_{12}^{\alpha,\beta,\gamma})\ :\ \ C = \left(
\begin{array}{cccccc}
 1 & 0 & 0 & 0 & 0 & 0 \\
 0 & 1 & 0 & 0 & 0 & 0 \\
 0 & 0 & 1 & 0 & 0 & 0 \\
 0 & 0 & 0 & 1 & 0 & 0 \\
 0 & \frac{\alpha }{2}-\frac{\beta }{2}+\frac{\gamma }{4} & \frac{\beta }{2}-\frac{\gamma }{4} & 0 & 1 & 0 \\
 0 & \frac{\beta }{2}-\frac{\gamma }{4} & \frac{\gamma }{2} & 0 & 0 & 1
\end{array}
\right)\]

\[(\cc|\Ajd) \rightarrow (C^2_{-1}|N_{12}^{\alpha,\kappa,\gamma})\ :\ \ C = \left(
\begin{array}{cccccc}
1 & 0 & 0 & 0 & 0 & 0 \\
 0 & -\frac{1}{\kappa } & 0 & 0 & 0 & 0 \\
 0 & 0 & 0 & 0 & 0 & 1 \\
 0 & 0 & 0 & 1 & 0 & 0 \\
 0 & -\frac{\alpha }{2 \kappa } & 0 & 0 & -\kappa  & 0 \\
 0 & 0 & -1 & 0 & 0 & -\frac{\gamma }{2}
\end{array}
\right), \kappa\neq 0\] ${\mathbf{ DD_{(2,4)} V_p,\ p\neq 0}}$ :
\[(\cpp|\Ajd) \rightarrow (\cpp|N_{12}^{\alpha,\beta,\gamma}) \ :\ \ C = \left(
\begin{array}{cccccc}
 1 & 0 & 0 & 0 & 0 & 0 \\
 0 & 1 & 0 & 0 & 0 & 0 \\
 0 & 0 & 1 & 0 & 0 & 0 \\
 0 & 0 & 0 & 1 & 0 & 0 \\
 0 & \frac{2 \alpha  p^2-2 \beta  p+\alpha +\gamma }{4 \left(p^3+p\right)} & \frac{\alpha +2 p \beta -\gamma }{4 p^2+4} & 0 & 1 & 0 \\
 0 & \frac{\alpha +2 p \beta -\gamma }{4 p^2+4} & \frac{2 \gamma  p^2+2 \beta  p+\alpha +\gamma }{4 \left(p^3+p\right)} & 0 & 0 & 1
\end{array}
\right)\] ${\mathbf{ DD_{(2,4)} V_0}}$ :
\[
(C^5_0|\Ajd) \rightarrow (C^5_0|N_{12}^{\alpha,\beta,-\alpha}) \ :\
\ C = \left(
\begin{array}{cccccc}
1 & 0 & 0 & 0 & 0 & 0 \\
 0 & 1 & 0 & 0 & 0 & 0 \\
 0 & 0 & 1 & 0 & 0 & 0 \\
 0 & 0 & 0 & 1 & 0 & 0 \\
 0 & 0 & \frac{\alpha }{2} & 0 & 1 & 0 \\
 0 & \frac{\alpha }{2} & \beta  & 0 & 0 & 1
\end{array}
\right)
\]
${\mathbf{ DD_{(2,4)} VI}}$ :\[ (\cdn|N_{12}^{0,0,1})  \rightarrow
(\cdn|N_{12}^{\alpha,\beta,\gamma})\ :\ \ C = \left(
\begin{array}{cccccc}
 1 & 0 & 0 & 0 & 0 & 0 \\
 0 & 1 & 0 & 0 & 0 & 0 \\
 0 & 0 & -\frac{1}{\sqrt{\gamma }} & 0 & 0 & 0 \\
 0 & 0 & 0 & 1 & 0 & 0 \\
 0 & \frac{\alpha }{2} & -\frac{\beta }{\sqrt{\gamma }} & 0 & 1 & 0 \\
 0 & \beta  & 0 & 0 & 0 & -\sqrt{\gamma }
\end{array}
\right), \ \gamma>0\]

\[ (\cdn|N_{12}^{0,0,1})  \rightarrow
(\cdn|N_{12}^{\alpha,\beta,\gamma})\ :\ \ C = \left(
\begin{array}{cccccc}
 -1 & 0 & 0 & 0 & 0 & 0 \\
 0 & 0 & 0 & 0 & 1 & 0 \\
 0 & 0 & \frac{1}{\sqrt{-\gamma }} & 0 & 0 & 0 \\
 0 & 0 & 0 & -1 & 0 & 0 \\
 0 & -1 & \frac{ \beta }{\sqrt{-\gamma }} & 0 & \frac{\alpha }{2} & 0 \\
 0 & 0 & 0 & 0 & \beta  &  \sqrt{-\gamma }
\end{array}
\right), \ \gamma<0\] ${\mathbf{ DD_{(2,4)} VII}}$ :\[
(\ct|N_{12}^{0,0,1}) \rightarrow (\ct|N_{12}^{\alpha\beta,\gamma}) \
:\ \ C =  \left(
\begin{array}{cccccc}
 \gamma  & 0 & 0 & 0 & 0 & 0 \\
 0 & \gamma  & 0 & 0 & 0 & 0 \\
 0 & 0 & 1 & 0 & 0 & 0 \\
 0 & 0 & 0 & \frac{1}{\gamma } & 0 & 0 \\
 0 & 0 & \frac{\alpha }{2} & 0 & \frac{1}{\gamma} & 0 \\
 0 & \frac{\alpha  \gamma }{2} & \beta  & 0 & 0 & 1
\end{array}
\right), \ \gamma\neq 0\] ${\mathbf{ DD_{(2,4)} VIII}}$ :\[
(C^5_0|N_{12}^{0,0,1}) \rightarrow
(C^5_0|N_{12}^{\alpha,\beta,\gamma}) \ :\ \ C = \left(
\begin{array}{cccccc}
 \epsilon & 0 & 0 & 0 & 0 & 0 \\
 0 & -\frac{1}{\sqrt{|\alpha +\gamma| }} & 0 & 0 & 0 & 0 \\
 0 & 0 & -\epsilon\frac{1}{\sqrt{|\alpha +\gamma| }} & 0 & 0 & 0 \\
 0 & 0 & 0 & \epsilon & 0 & 0 \\
 0 & \frac{\beta }{\sqrt{|\alpha +\gamma| }} & -\epsilon\frac{\alpha }{2 \sqrt{|\alpha +\gamma| }} & 0 & -\sqrt{|\alpha +\gamma| } & 0 \\
 0 & -\frac{\alpha }{2 \sqrt{|\alpha +\gamma| }} & 0 & 0 & 0 & -\epsilon\sqrt{|\alpha +\gamma| }
\end{array}
\right),
\] $$\hskip 6cm \alpha +\gamma\neq 0,\ \epsilon=sign(\alpha +\gamma)$$


\begin{thebibliography}{1}

\bibitem{leiserg:solcyb}D. Leites and V. Serganova  { Solutions of the
classical Yang--Baxter equation for simple Lie superalgebras}.
{\small\it Theoret. Mat. Fiz} {\bf 58}, 26 (1984).

\bibitem{karaa:rmat} G. Karaali,  {Constructing $r$--matrices on
simple Lie superalgebras}. {\small\it J. Algebra} {\bf 282} 83
(2004).

\bibitem{klse:dna} C.Klim\v c\'{\i}k and P.\v Severa.
 { Dual non--{A}belian duality and the {D}rinfeld double. }
{\small\it Phys.Lett. B} {\bf 351}, 455 (1995).

\bibitem{unge:pltp} R. von Unge,  {Poisson--Lie T--plurality}, {\small\it J. High
Energy Phys.} 02:07(2002)014.

\bibitem{jusob:e2osp12}C. Juszczak and J.T. Sobczyk, {  Classification of low dimensional Lie
super--bialgebras}. {\small\it J. Math.Phys.} {\bf 19} 2400 (1998).

\bibitem{ehra:cfn23sba} A.Eghbali, F.Heidapour and
A.Rezaei-Aghdam.  { Classification of two and three dimensional Lie
super-bialgebras.} arXiv:0901.4471 [math-ph].

\bibitem{backh:class} N.Backhouse.
A classification of four--dimensional Lie superalgebras. {\small\it
J. Math. Phys.} {\bf 19}, 2400 (1978).

\bibitem{gom:ctd} X. Gomez
 Classification of three--dimensional Lie bialgebras.
{\small\it  J. Math. Phys.} {\bf  41}, 4939 (2000).

\bibitem{hlasno:pltdm2dt} L.~Hlavat\'y and L.~\v{S}nobl.
Poisson--{L}ie {T}--dual models with two--dimensional targets.
{\small\it  Mod. Phys. Lett. A}  {\bf 17}, 429 (2002).

\bibitem{snohla:ddoubles}
L.~\v{S}nobl and L.~Hlavat\'y,  {Classification of 6-dimensional
  real Drinfel'd doubles}, {\small\it Int.J.Mod.Phys. A} {\bf 17}, 4043 (2002).

\bibitem{ro:supman} A. Rogers, {  Supermanifolds, Theory and
Applications}, World Scientific, Singapore, 2007.

\bibitem{andru:lspls}
 N. Andruskiewitsch,
 Lie Superbialgebras and Poisson--Lie Supergroups.
{\small\it  Abh. Math. Sem. Univ. Hamburg} {\bf 63}, 147 (1993).


\bibitem{olshan} G.I. Olshanski, { Quantized universal enveloping superalgebra of type
$Q$}, {\small\it Lett.Math.Phys. {\bf 24}, 93 (1992)}.

\end{thebibliography}
\end{document}